\documentclass[sn-mathphys,Numbered]{sn-jnl}

\usepackage{graphicx}
\usepackage{multirow}
\usepackage{amsmath,amssymb,amsfonts}
\usepackage{amsthm}
\usepackage{mathrsfs}
\usepackage[title]{appendix}
\usepackage{xcolor}
\usepackage{textcomp}
\usepackage{manyfoot}
\usepackage{booktabs}
\usepackage{algorithm}
\usepackage{algorithmicx}
\usepackage{algpseudocode}
\usepackage{listings}
\usepackage[version=3]{mhchem}
\usepackage{siunitx}
\usepackage{comment}
\usepackage{diagbox}
\usepackage{mathtools}
\usepackage[hang,small,bf]{caption}
\usepackage[subrefformat=parens]{subcaption}
\newtheorem{theorem}{Theorem}
\newtheorem{proposition}[theorem]{Proposition}

\newtheorem{remark}{Remark}

\begin{document}

\title[Article Title]{Analysis of Signal Distortion in Molecular Communication Channels Using Frequency Response}

\author[1]{\fnm{Shoichiro} \sur{Kitada}}\email{kitada.shoichiro@keio.jp}
\equalcont{These authors contributed equally to this work.}

\author[1]{\fnm{Taishi} \sur{Kotsuka}}\email{tkotsuka@keio.jp}
\equalcont{These authors contributed equally to this work.}

\author*[1]{\fnm{Yutaka} \sur{Hori}}\email{yhori@appi.keio.ac.jp}

\affil*[1]{\orgdiv{Department of Applied Physics and Physico-Informatics}, \orgname{Keio University}, \orgaddress{\street{3-14-1}, \city{Yokohama},  \state{Kanagawa} \postcode{223-8522}, \country{Japan}}}

\abstract{Molecular communication (MC) is a concept in communication engineering, where diffusive molecules are used to transmit information between nano or micro-scale chemical reaction systems. Engineering MC to control the reaction systems in cells is expected for many applications such as targeted drug delivery and biocomputing.  
Toward control of the reaction systems as desired via MC, it is important to transmit signals without distortion by MC since the reaction systems are often triggered depending on the concentration of signaling molecules arriving at the cells. 
In this paper, we propose a method to analyze signal distortion caused by diffusion-based MC channels using frequency response of channels. The proposed method provides indices that quantitatively evaluate the magnitude of distortion and shows parameter conditions of MC channels that suppress signal distortion. Using the proposed method, we demonstrate the design procedure of specific MC channels that satisfy given specifications. Finally, the roles of MC channels in nature are discussed from the perspective of signal distortion.}


\maketitle

\section{Introduction}\label{sec1}
In nature, many cells share information using signaling molecules to achieve cooperative functions \cite{bassler2002small,waters2005quorum,armingol2021deciphering}. This information-sharing system between cells is known as molecular communication (MC) systems. In MC systems, a transmitter cell releases signaling molecules, and a receiver cell accepts the signal by adsorption or absorption of the signaling molecules, triggering chemical reactions in the cell. One of many examples of MC is quorum sensing conducted by bacteria, where the cells use signaling molecules so called autoinducer to regulate population density dependent collective behaviors \cite{de2000bacterial,miller2001quorum}. 
MC systems are also engineered in synthetic biology to perform complex tasks at a population level. Potential applications of such synthetic MC systems include targeted drug delivery and biocomputing \cite{femminella2015molecular,suda2018molecular,Martins2022,Bi2021,kuzuya2023}. 
To control systems via MC for these applications, MC channels need to be able to transfer information appropriately, which inspires us to analyze fundamental characteristics of MC channels.

\par 

\begin{figure}[tb]
    \centering
    \includegraphics[width=0.9\textwidth]{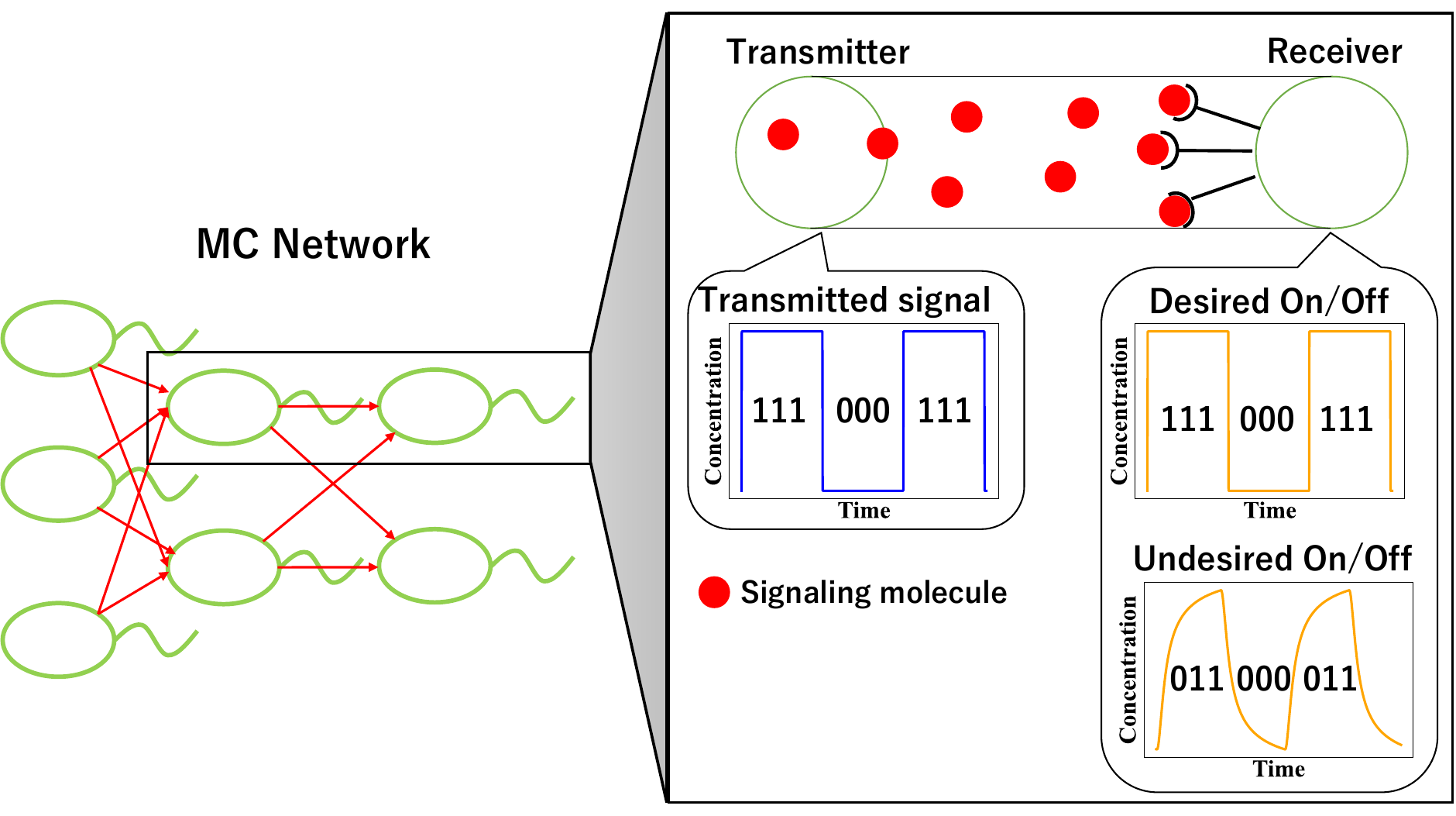}
    \caption{Network of multiple cells and MC between a transmitter cell and a receiver cell with desired and undesired on/off signal profiles at the receiver.}
    \label{fig:mc_network}
\end{figure}

In the control of reaction systems in a receiver cell, MC channels must be designed so that they can transfer desired signals from a transmitter cell to a receiver cell. To this end, the gain characteristics of MC channels were analyzed, and the frequency bandwidth that can be transmitted without attenuating the magnitude of the signal was revealed \cite{pierobon2010physical, chude2015diffusion, kotsuka2022frequency, kotsuka2023}. However, the analysis of the simple gain characteristics does not necessarily assess the ability of the transmission signals to trigger the downstream reactions in the receiver cells since the regulation of reaction systems in a cell often involves a digital-like switching mechanism modeled by the Hill function \cite{hill1910possible}. 
In other words, the signal may not be able to transfer the desired on/off patterns to the reaction systems at a receiver cell if the waveform of the signal at the receiver is distorted as shown in Fig. \ref{fig:mc_network}. Therefore, to guarantee the quality of the transmitted signals, it is crucial to assess the degree of distortion that is added in MC channels. 

\par
In this paper, we propose a method to analyze signal distortion caused by one-dimensional diffusion-based MC channels. Specifically, we first introduce indices evaluating signal distortion by the gain and the phase delay characteristics and derive these characteristics of MC channels based on the diffusion equation and the rate equation. We then show design conditions for MC channels in which the magnitude of distortion becomes below a specified level based on the indices. Using the proposed method, we demonstrate the design procedure of specific MC channels that satisfy given specifications. Finally, the roles of MC channels in nature are discussed from the perspective of signal distortion.

\par
This paper is organized as follows. In the next section, we model one-dimensional MC channels based on diffusion of signaling molecules and binding/dissociation reactions between signaling molecules and receptors. In Section 3, we define indices evaluating signal distortion based on the frequency response and analyze the magnitude of distortion in the channels quantitatively and show the design condition of MC channels to repress the signal distortion under certain level. Using the analytical results, we demonstrate a design procedure of MC channels that can suppress the signal distortion in Section 4. Finally, the paper is concluded in Section 5.\par

\section{Model of MC Channels and Problem Setting}
\begin{figure}[t]
\begin{center}
\includegraphics[width=0.9\textwidth]{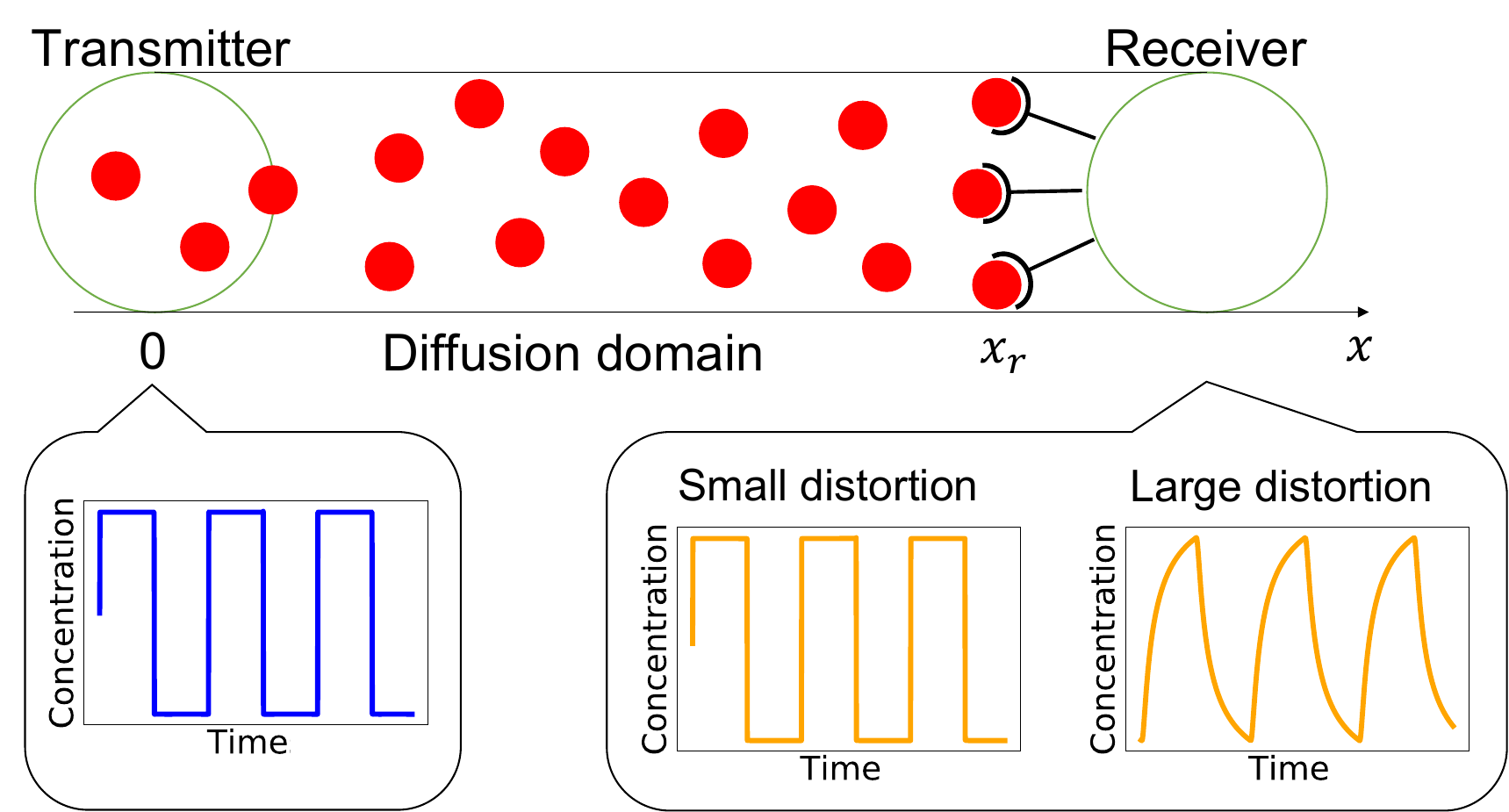}
\caption{One-dimensional model of MC channel, where a pair of a transmitter and a receiver cell is located, and illustration of} signals with distortion
\label{fig:mcchannel}
\end{center}
\end{figure}

We consider one-dimensional MC channels in which signaling molecules secreted from a transmitter cell diffuse in the fluidic environment $\Omega = [0,\infty)$ and their concentration is detected by a receiver cell at $x=x_r$ as shown in Fig. \ref{fig:mcchannel}, where $x\in\Omega$ is the position. The MC channels can be decomposed into the diffusion system in which the signaling molecules diffuse to the position of the receiver cell and the reception system in which they are received by receptors on the receiver cell.\par
In the diffusion system, the signaling molecules move in a fluidic medium according to the following diffusion equation,

\begin{equation} 
\frac{\partial u(x,t)}{\partial t} = \mu\frac{\partial^2 u(x,t)}{\partial x^2},
\label{eq:diffusion}
\end{equation}
where $u(x,t)$ is the concentration of the signaling molecules at position $x$ and at time $t$, and $\mu$ is the diffusion coefficient of the signaling molecules. At the position of the receiver cell $x=x_r$, the concentration of signaling molecules $u(x_r,t)$ is sensed by binding/unbinding of the signaling molecule to the receptor. 
We assume that the concentration of the signaling molecules $u(x,t)$ does not change due to the binding and the dissociation of the molecule to the receptor since, in typical receptor-ligand reactions, the rate of dissociation is larger than that of binding \cite{cozens1990receptor}, resulting in the immediate dissociation of the signaling molecule from the receptor.
Since signaling molecules emitted at $x=0$ diffuse to infinity beyond the position of the receiver cell $x=x_r$, the boundary conditions are
\begin{align}
    &u(0,t) = v(t),\\
    &\lim_{x \to \infty}u(x,t) = 0,
    \label{eq:boundary1}
\end{align}
where $v(t)$ is the concentration of the emitted molecules.

\begin{figure}[t]
    \centering    \includegraphics[width=0.6\textwidth]{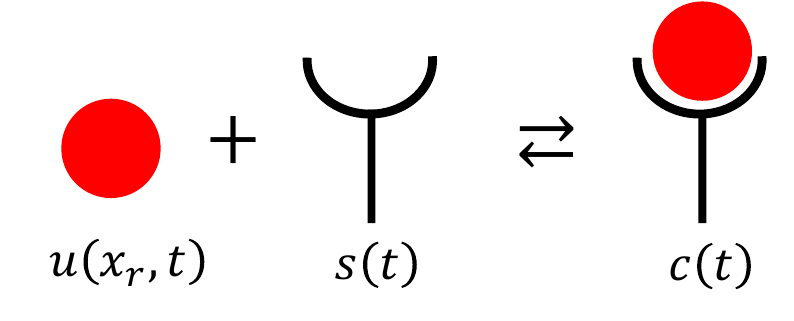}
    \caption{Schematic diagram of binding and dissociation reaction between signaling molecules and receptors}
    \label{fig:reception}
\end{figure}

Next, we consider the reception system, where signaling molecules arriving at position $x = x_r$ bind to receptors on the surface of the receiver cell and form a complex or dissociate from the complex. A schematic diagram of this process is shown in Fig. \ref{fig:reception}. Let $k_f$ and $k_r$ denote the rate constant in the binding reaction of the signaling molecule to the receptor and the rate constant in the dissociation reaction from the complex, respectively. The binding and dissociation reactions between the signaling molecules and the receptors can be expressed as
\begin{equation}
    \ce{{\it s(t) + u(x_r, t)} <=>[k_f][k_r] {\it c(t)}},
    \label{eq:chem}
\end{equation}
where $c(t)$ is the concentration of the complex, and $s(t)$ is the concentration of the free receptors. 
Assuming $k_{f}u(x_r,t)\ll k_r$, the dynamics of $c(t)$ can be expressed as 
\begin{equation}
   \frac{dc(t)}{dt}=k_{f}ru(x_r,t)-k_{r}c(t),
   \label{eq:rateeq3} 
\end{equation}
where $r=s(t)+c(t)$ is the total concentration of the receptor on the cell \cite{shahmohammadian2013nano,lauffenburger1996receptors}. Note that since the signaling molecules keep diffusing in MC channel, the concentration of signaling molecules $u(x_r,t)$ at the receiver cell would be sufficiently smaller than the ratio $k_r/k_f$ \cite{cozens1990receptor}, which leads to $k_{f}u(x_r,t)\ll k_r$.
\par

We consider the concentration waveform of a signaling molecule at $x=0$ as an input signal, which is transmitted to $x=x_r$ by diffusion. To transfer the desired signal to the reaction system in the receiver cell, the input concentration waveform should not be distorted as shown in Fig. \ref{fig:mcchannel}. In particular, the magnitude of the distortion caused by the diffusion system should be smaller than that caused by the reception system to keep the distortion of the entire MC channel small. Therefore, we quantitatively analyze the magnitude of distortion caused by signal transmission by diffusion to provide design guidelines for MC channels to suppress signal distortion.

\section{Characteristics of Distortion in MC Channels}
In this section, we propose a method to quantitatively analyze the magnitude of distortion caused by the diffusion and the reception system. To this end, we first define the magnitude of signal distortion by two types of features based on frequency response. We then analyze the magnitude of the distortion based on the transfer functions of the diffusion systems and the reception system. Finally, we show the parameter conditions of the diffusion systems for the magnitude of the distortion to be below a predefined threshold.
\subsection{Amplitude distortion and delay distortion}
We consider a transfer function $F(s)$ representing a general linear system. When we input sine wave $v(t)=\sin(\omega t)$ into the system $F(s)$, the output signal $z(t)$ is expressed as
\begin{eqnarray}
      z(t) &= |F(j\omega)|\sin(\omega t+\angle F(j\omega))\\
      &= |F(j\omega)|\sin(\omega (t-\tau(\omega))),
\end{eqnarray}
where 
\begin{equation}
    \tau(\omega)=-\frac{\angle F(j\omega)}{\omega}
\end{equation}
is the phase delay, which represents the delay time of the system \cite{lathi2005linear}. If the gain $|F(j\omega)|$ and the phase delay $\tau(\omega)$ vary depending on frequency $\omega$, the output waveform is distorted. Distortion due to the gain $|F(j\omega)|$ is called amplitude distortion, and distortion due to the phase delay $\tau(\omega)$ is called delay distortion \cite{carlson1968communication,DiToro1948}. To quantitatively analyze the magnitude of distortion on frequency basis, we decompose distortion into amplitude distortion and delay distortion and define their respective indices.
\par
We define the index representing the magnitude of amplitude distortion in the frequency bandwidth $[\omega_1, \omega_2]$ as
\begin{equation}
 Q:=\max_{\omega_1\leq\omega\leq\omega_2}g(\omega)-\min_{\omega_1\leq\omega\leq\omega_2}g(\omega),
 \label{eq:Q}
\end{equation}
where
\begin{equation}
    g(\omega) = 20\log{|F(j\omega)|}
\end{equation}
is the logarithmic gain characteristics of the frequency transfer function $F(j\omega)$. The amplitude distortion $Q$ represents the largest difference between the logarithmic gain characteristics of the two sine waves which have different frequencies in $[\omega_1, \omega_2]$. We use the logarithmic gain characteristics in the index $Q$ since it is relatively easy to compute the logarithm of the gain characteristics of the diffusion systems, which is discussed in detail in Sec. 3.2. In the same way, we define the index representing the magnitude of delay distortion in the frequency bandwidth $[\omega_1, \omega_2]$ as
\begin{equation}
 R:=\frac{1}{T_1}\left(\max_{\omega_1\leq\omega\leq\omega_2}{\tau(\omega)}-\min_{\omega_1\leq\omega\leq\omega_2}{\tau(\omega)}\right),
 \label{eq:R}
\end{equation}
where $\tau(\omega)$ is the phase delay of the signal with the frequency $\omega$ and $T_1$ is the period of the signal with $\omega_1$, namely $T_1 = 2\pi/\omega_1$. The delay distortion $R$ expresses the largest difference between the phase delays of the two sine waves, which have different frequencies in $[\omega_1, \omega_2]$, normalized by the period $T_1$. Since the smaller the difference in gain (phase delay, resp.) in the frequency bandwidth of the signal, the smaller the magnitude of distortion due to amplitude shift (phase delay shift, resp.), the indices $Q$ ($R$, resp.) are smaller if the system can transfer signals with less distortion. Therefore, we evaluate the magnitude of distortion caused by MC channels using the indices $Q$ and $R$. To obtain the amplitude distortion $Q$ and the delay distortion $R$ in MC channels, we need the logarithmic gain characteristics $g(\omega)$ and the phase delay $\tau(\omega)$ of the diffusion system and the reception system. To this end, we derive the transfer functions and the indices of the diffusion system and the reception system in the following sections. To quantitatively analyze the magnitude of distortion caused by diffusion, we first define the magnitude of signal distortion by two types of features based on frequency characteristics. We then analyze the magnitude of the distortion based on the transfer function of the diffusion system and show the parameter conditions for the distortion to be below a predefined threshold.

\subsection{Transfer function and signal distortion of the diffusion system }

\par
\smallskip
To obtain the amplitude distortion $Q$ and the delay distortion $R$ for the diffusion system, we derive the transfer function from the diffusion equation (\ref{eq:diffusion}). Using the Laplace transform of the concentrations $u(x_r,t)$ and $u(0,t)$ for time $t$ and considering their ratio, we obtain 
\begin{equation}
    G(s) \coloneqq \frac{U(x_r,s)}{U(0,s)}=e^{-\sqrt{\frac{x_r^{2}}{\mu}s}},
    \label{eq:difsys}
\end{equation}
where $U(\cdot, s)$ is the Laplace transform of the concentration $u(\cdot, t)$. Using Eq. (\ref{eq:difsys}), we have the following proposition, which gives an analytic form of the amplitude and delay distortion, $Q$ and $R$, for a frequency bandwidth $[\omega_1, \omega_2]$. 

\noindent
\begin{proposition}
    Consider the diffusion system $G(s)$ in Eq. (\ref{eq:difsys}). The amplitude distortion $Q_G$ in Eq. (\ref{eq:Q}) and the delay distortion $R_G$ in Eq. (\ref{eq:R}) for the diffusion system $G(s)$ are 
  \begin{equation}
  Q_G = g_G(\omega_1) - g_G(\omega_2) = 20\sqrt{\frac{x_r^{2}}{2\mu}}(\sqrt{\omega_2}-\sqrt{\omega_1})\log_{10}{e},
  \label{eq:Qd}
  \end{equation}
  \begin{equation}
  R_G = \frac{\tau_G(\omega_1)-\tau_G(\omega_2)}{T_1} =\frac{1}{T_1}\sqrt{\frac{x_r^2}{2\mu}}\left(\frac{1}{\sqrt{\omega_1}}-\frac{1}{\sqrt{\omega_2}}\right),
  \label{eq:Rd}
  \end{equation}
  where $g_G(\cdot)$ and $\tau_G(\cdot)$ are the gain characteristics and the phase delay of the diffusion system $G(j\omega)$, respectively.
\end{proposition} 

\par
\smallskip
The proof of Proposition 1 is shown in Appendix \ref{sec:prop1}. Equations (\ref{eq:Qd}) and (\ref{eq:Rd}) indicate that the magnitude of distortion decreases as the communication distance $x_r$ decreases and as the diffusion coefficient $\mu$ increases. Furthermore, they represent that the magnitude of distortion increases as the signal bandwidth $[\omega_1, \omega_2]$ goes into the higher frequency region. Since the gain characteristics of the diffusion system involves an exponential function, taking the logarithm of the gain characteristics simplifies the calculation of the index $Q_G$. 

\medskip
\begin{remark}

We have considered one-dimensional MC channels $\Omega$ to present the essential idea of the analysis of signal distortion. The same idea can be extended to a more general and realistic case, where the molecules diffuse in three dimensions. 
In the case of diffusion in three dimensional space, the transfer function $G'(s)$ of the diffusion system from the transmitter cell to the receiver cell can be derived as
\begin{equation}
    G'(s) = \frac{1}{4\pi\mu d} e^{-\sqrt{\frac{d^{2}}{\mu}s}},
\end{equation}
where $d$ is the distance between the transmitter cell and the receiver cell \cite{schiff1999laplace, mosetti1985analysis}. The transfer function $G'(s)$ consists of the function $G(s)$ with the constant coefficient $1/4\pi\mu d$, which is independent of the frequency variable $s$. 
Therefore, the corresponding indices $Q_{G'}$ and $R_{G'}$ for the diffusion system in three dimensional space differ from $Q_{G}$ and $R_{G}$ only by the constant factor, which leads to the similar conclusion to the one-dimensional case.
\end{remark}

\subsection{Transfer function of the reception system}
Next, we derive the transfer function of the reception system to obtain the amplitude distortion $Q$ and the delay distortion $R$ for the reception system. Using the Laplace transform of both sides of the differential equation (\ref{eq:rateeq3}), the transfer function of the reception system $H(s)$ is obtained as
\begin{equation}
H(s) \coloneqq \frac{C(s)}{U(x_r,s)}= \frac{k_f r}{s + k_r},
\label{eq:H(s)}
\end{equation}
where $C(s)$ is the Laplace transform of $c(t)$. Using Eq. (\ref{eq:H(s)}), we have the following proposition, which gives an analytic form of the amplitude and delay distortion, $Q$ and $R$ for the reception system, for a frequency bandwidth $[\omega_1, \omega_2]$. 

\begin{proposition}
Consider the reception system $H(s)$ shown as Eq. (\ref{eq:H(s)}). The amplitude distortion $Q_H$ and delay distortion $R_H$ for the reception system are
\begin{equation}
    Q_H = g_H(\omega_1) -g_H (\omega_2) = 20\log_{10}{\frac{\sqrt{k_r^2+\omega_2^2}}{\sqrt{k_r^2+\omega_1^2}}},
\label{eq:Q_reception}
\end{equation}
\begin{equation}
    R_H = \tau_H(\omega_1)-\tau_H(\omega_2) = \frac{1}{2\pi}\left(\arctan\left(\frac{\omega_1}{k_r}\right)-\frac{\omega_1}{\omega_2}\arctan\left(\frac{\omega_2}{k_r}\right)\right),
\label{eq:R_reception}
\end{equation}
where $g_H(\omega)$ is the gain characteristics, and $\tau_H(\omega)$ is the phase delay characteristics for the frequency transfer function $H(j\omega)$.
\end{proposition}
The proof of Proposition 2 is shown in Appendix \ref{sec:prop2}. Equations (\ref{eq:Q_reception}) and (\ref{eq:R_reception}) indicate that the magnitude of distortion caused by the reception system increases as the signal bandwidth $[\omega_1, \omega_2]$ goes into the higher-frequency region as same as that caused by the diffusion system.

\par
\subsection{Upper limit of the communication distance satisfying distortion constraint}
Using the Proposition 1 and Proposition 2, we show the condition of the communication distance to repress signal distortion below a certain level based on the indices of the amplitude distortion and the delay distortion for the diffusion system and the reception system. Let $Q_{M}$ and $R_{M}$ denote the amplitude distortion and the phase distortion of the MC channel $G(s)H(s)$. Since the gain and phase characteristics for the diffusion system $G(s)$ and the reception system $H(s)$ are monotonically decreasing functions for frequency $\omega$, $Q_{M}$ and $R_{M}$ can be decomposed as
\begin{equation}
    Q_{M} = Q_{G} + Q_{H},
    \label{eq:Q_GiH}
\end{equation}
\begin{equation}
    R_{M} = R_{G} + R_{H}.
    \label{eq:R_GiH}
\end{equation}
Therefore, by setting arbitrary thresholds $Q_0> Q_M$ and $R_0> R_M$ for the amplitude distortion and the phase distortion of the MC channel $G(s)H(s)$, the upper limit on the communication distance $x_r$ between the cells to keep the magnitude of distortion below a threshold value is obtained as
\begin{equation}
    x_r < \min{\left(x_Q, x_R\right)},
\label{eq:design_condition_xr}
\end{equation}
where
\begin{equation}
x_Q := \frac{\sqrt{2\mu}(Q_0-Q_H)}{20(\sqrt{\omega_2}-\sqrt{\omega_1})\log_{10}e},
\label{eq:x_Q}
\end{equation}
\begin{equation}
x_R := \frac{\sqrt{2\mu}(R_0-R_H)T_1}{\left(\frac{1}{\sqrt{\omega_1}}-\frac{1}{\sqrt{\omega_2}}\right)}.
\label{eq:x_R}
\end{equation}
This can be an indicator to design the communication distance of MC channels.

\section{Numerical Simulation}
In this section, we show a design procedure for a specific MC channel that can transfer signals with small distortion. In particular, we use Proposition 1 and 2 to show the design condition of the communication distance $x_r$ under a given specification. We then analyze the effect by $x_r$ as well as other parameters to the distortion. Finally, we discuss the structure of MC channels in nature with a focus on signal distortion.

\subsection{Design of communication distance for suppressing distortion in MC channel}

We consider the MC channel in which signaling molecules such as autoinducer are secreted from a transmitter cell, diffuse through the fluidic environment, and are then sensed by receptors on the surface of a receiver cell. The concentration of the complex formed by signaling molecules and receptors on the receiver cell is observed when a square wave signal is input into this MC channel. The observed signal is distorted due to the kinetics of the diffusion of molecules and reception on the receiver cell. Hence, to suppress the distortion, we here consider adjusting spatial arrangement of cells for given transmitter/receiver cells, and signaling molecules. The specific objective of the design is to find a communication distance $x_r$ such that the distortion due to the diffusion system becomes less than $1/5$ times the distortion due to the reception system.
Thus, the thresholds for the magnitude of distortion are set as $Q_0=1.2Q_H$ and $R_0=1.2R_H$. 
Specifications of the MC channel are as follows:
\begin{itemize}
    \item [(a)] The diffusion coefficient of the signaling molecules is $\mu=83\,\si{\micro m^2/s}$ \cite{karig2018stochastic}.
    \item [(b)] The design range of the communication distance $x_r$ is limited by $0\leq x_r \leq 400\,\si{\micro m}$.
    \item [(c)] Each parameter of the reception system is $k_r=4.0\cdot10^{-3}\,\si{s^{-1}}$, $k_f=1.0\cdot10^{-3}\,\si{\micro M^{-1}s^{-1}}$ and $r=4.0\,\si{\micro M}$ \cite{lauffenburger1996receptors}.
\end{itemize}
\par
To design the communication distance $x_r$ under the specifications (a)--(c), we first analyze the magnitude of the distortion based on the indices of the amplitude distortion in Eq. (\ref{eq:Q}) and the delay distortion in Eq. (\ref{eq:R}) for the diffusion system and the reception system. We then evaluate the distortion of signals with the frequency band $[5.0\cdot10^{-4}, 4.0\cdot10^{-1}]\,\si{rad/s}$, where the upper bound is the frequency at which the signal transmission is limited by the receptor system rather than diffusion since the gain becomes $|H(j\omega_2)|=1/100$. The magnitudes of the distortions of the reception system in this frequency band are computed as $Q_H=39.9$ and $R_H=1.95\cdot10^{-2}$ using the parameters in the specification (c). Substituting the values of $Q_H$ and $R_H$ into the formula (\ref{eq:design_condition_xr}), we obtain the condition for the communication distance as $x_r < 14.6\,\si{\micro m}$. If $x_r$ satisfies this condition, the magnitude of distortion due to the diffusion system $G(s)$ can be suppressed to less than $1/5$ times that of distortion in the reception system $H(s)$. \par
\begin{figure}[t]
    \centering
    \includegraphics[width=0.7\textwidth]{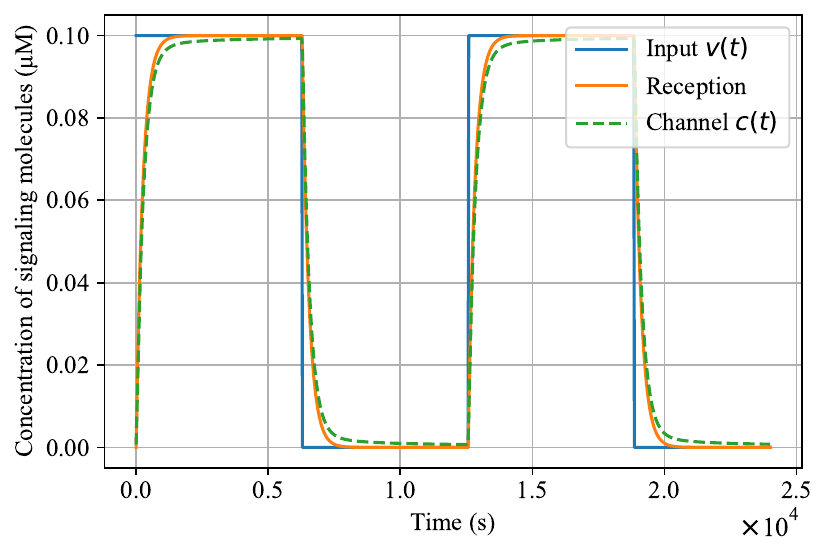}
    \caption{The responses of the reception system and the MC channel to a square wave with the communication distance $x_r=14\,\si{\micro m}$, which satisfies the design condition (\ref{eq:design_condition_xr}). The distortion level of the signal transmitted via both of the diffusion and the reception system (green) is almost the same as that of the signal transmitted only via the reception system (orange), indicating that the primary source of the distortion is the reception system. }
    \label{fig:diffusion_infinite_distortion_small}
\end{figure}
\begin{figure}[t]
    \centering
    \includegraphics[width=0.7\textwidth]{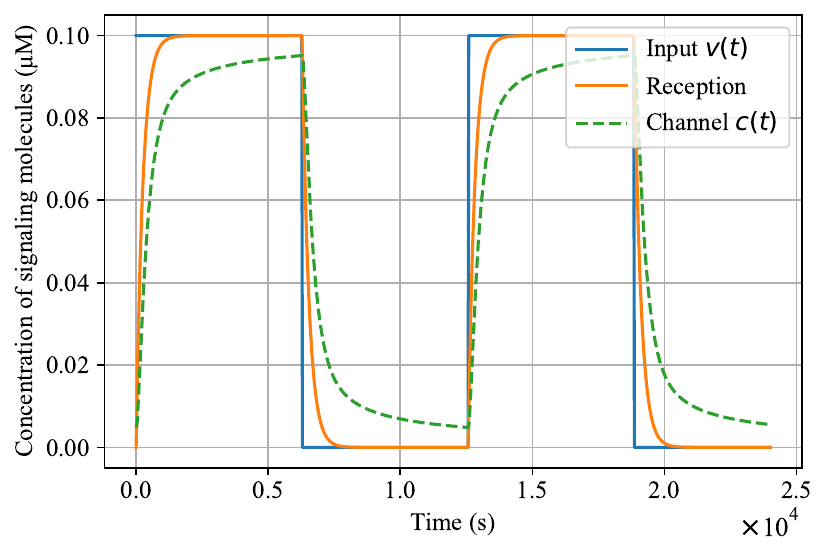}
    \caption{The responses of the reception system and the MC channel to square wave with the communication distance $x_r=100\,\si{\micro m}$, which does not satisfy the design condition (\ref{eq:design_condition_xr}). Compared to Fig. \ref{fig:diffusion_infinite_distortion_small}, the signal is further distorted by the diffusion system.}
    \label{fig:diffusion_infinite_distortion_large}
\end{figure}

Fig. \ref{fig:diffusion_infinite_distortion_small} illustrates that the responses in the MC channel $G(s)H(s)$ (green dotted line) and the reception system $H(s)$ (orange line) with the communication distance $x_r=14\,\si{\micro m}$ when the square wave $v(t)$ with the fundamental frequency $\omega_1=5.0\cdot10^{-4}\,\mathrm{rad/s}$ is input to the systems. Since the distortion in the diffusion system is suppressed more than that in the reception system, the distortion in the channel is approximately at the same level as the reception system. On the other hand, Fig. \ref{fig:diffusion_infinite_distortion_large} illustrates the responses in the MC channel $G(s)H(s)$ (green dotted line) and the reception system $H(s)$ (orange line) with the communication distance $x_r=100\,\si{\micro m}$. Fig. \ref{fig:diffusion_infinite_distortion_large} shows that the distortion of the response in the MC channel is more significant than that in the reception system alone since the rise of the response in the MC channel is further slowed down due to the effect of the diffusion of signaling molecules. These examples illustrate that one can design the MC channel satisfying the specifications (a) -- (c) based on the design condition \eqref{eq:design_condition_xr} derived from the proposed indices $Q$ and $R$. Note that the proposed method can be applied to signals composed of arbitrary frequency bands. This feature allows for the analysis of signal distortion for ion-based communication as well, where the signal tends to be an impulse-like shape that can be modeled as step signals with small width. \par
The distortion analysis is particularly useful to ensure the activation timing of the downstream reaction in the receiver cell. Suppose the reaction system in the receiver cell is activated when the concentration of the complex is $c(t)\geq0.09\,\si{\micro M}$. We define the desired activation time $T_v$ as the time when the signal becomes $c(t)\geq0.09\,\si{\micro M}$ within one pulse, and since the lowest frequency of the signal is $\omega_1 = 5.0\cdot10^{-4}\,\si{rad/s}$, the desired activation time is calculated as $T_v=6.5\cdot10^{3}\,\si{s}$. On the other hand, the activation time in the response of the MC channel response is $T_a=5.6\cdot10^{3}\,\si{s}$ for $x_r=19\,\si{\micro m}$ (Fig. \ref{fig:diffusion_infinite_distortion_small}) and $T_a=4.1\cdot10^{3}\,\si{s}$ for $x_r=100\,\si{\micro m}$ (Fig. \ref{fig:diffusion_infinite_distortion_large}). The activation time $T_a$ in the MC channel for $x_r=19\,\si{\micro m}$ is closer to the desired activation time $T_v$. This is because $x_r=19\,\si{\micro m}$ satisfies the design condition (\ref{eq:design_condition_xr}) and the distortion is smaller. Therefore, by designing MC channels using the proposed method, it is possible to approach the desired switching control for the reaction system in the receiver cell. \par

\par

\subsection{Dependence of distortions on parameters}

Next, we explore the dependence of the proposed indices on parameters such as the diffusion coefficient $\mu$ and the dissociation rate constant $k_r$. These parameters are usually specific to the species of the receiver cell and the signaling molecules. Thus, understanding the effect of these parameters on distortion allows for the proper choice of the cell and the molecule.

\par
\smallskip

To simplify the analysis of the relation between the distortion indices $Q_G$, $R_G$, $Q_H$, and $R_H$ 
and the parameters, we normalize the frequencies by
$\omega^{\prime}_1\coloneqq\omega_1/k_r$ and $\omega^{\prime}_2\coloneqq\omega_2/k_r$,
and the communication distance by 
\begin{equation}
    \lambda \coloneqq \sqrt{\frac{x_r^{2}k_r}{2\mu}}.
    \label{eq:lambda}
\end{equation}
Therefore, the distortion indices for the diffusion system and the reception system can be rewritten as 
\begin{equation}
    Q_G = 20\lambda\left(\sqrt{\omega^{\prime}_2}-\sqrt{\omega^{\prime}_1}\right)\log_{10}e,
    \label{eq:Q_G}
\end{equation}
\begin{equation}
    R_G = \frac{\omega^{\prime}_1}{2\pi}\lambda\left(\frac{1}{\sqrt{\omega^{\prime}_1}}-\frac{1}{\sqrt{\omega^{\prime}_2}}\right),
    \label{eq:R_G}
\end{equation}
\begin{equation}
    Q_H = 20\log_{10}\frac{\sqrt{1+{\omega^{\prime}_2}^{2}}}{\sqrt{1+{\omega^{\prime}_1}^{2}}},
    \label{eq:Q_H}
\end{equation}
\begin{equation}
    R_H = \frac{1}{2\pi}\left(\arctan{\omega^{\prime}_1}-\frac{\omega^{\prime}_1}{\omega^{\prime}_2}\arctan{\omega^{\prime}_2}\right),
    \label{eq:R_H}
\end{equation}
respectively. Equations (\ref{eq:Q_G})--(\ref{eq:R_H}) imply that the distortion indices for the diffusion system and the reception system only depend on three parameters: the normalized frequencies $\omega'_1$, $\omega'_2$, and the normalized distance $\lambda$. It should be noted that the distortion indices $Q_G$ and $R_G$ for the diffusion system is linear with respect to the normalized distance $\lambda$, indicating that the distortion is linearly proportional to the communication distance.

\par
\smallskip

Fig. 6 shows the distortion indices $Q_G$ and $R_G$ of the diffusion system for $\lambda=1$ and the indices $Q_H$ and $R_H$ of the reception system when the normalized frequencies is  varied betwee $0$ -- $100\,\si{rad}$. Fig. 6 indicates that the amplitude distortion $Q_G$ and $Q_H$ monotonically decrease with respect to the lowest frequency $\omega'_1$, and monotonically increase with respect to the highest frequency $\omega'_2$. On the other hand, the delay distortion $R_G$ and $R_H$ increase monotonically only for the highest frequency $\omega'_2$, but it takes a maximum value at some $\omega'_1$. Specifically, $R_G$ is maximum at $\omega'_1=\omega'_2/4$ and $R_H$ is maximum at $\omega'_1=\sqrt{-1+\omega'_2/\arctan\omega'_2}$, which can be derived by a simple partial differentiation of Eq. (\ref{eq:R_G}) and Eq. (\ref{eq:R_H}) with respect to $\omega'_1$. Therefore, narrowing the frequency bandwidth $[\omega'_1, \omega'_2]$ of the input signal by increasing the lowest frequency $\omega'_1$ does not necessarily result in less distortion. These results imply that, to design the MC channel such that the distortion due to the diffusion system is less than the distortion due to the reception system, we should first decide the normalized frequency bandwidth $[\omega'_1, \omega'_2]$ and then scale $Q_G$ and $R_G$ to be smaller than $Q_H$ and $R_H$ by tuning the normalized distance $\lambda$.

\begin{figure}[tb]
    \begin{tabular}{cc}
      \begin{minipage}[t]{0.45\hsize}
        \centering
        \subcaption{}
        \includegraphics[keepaspectratio, scale=0.4]{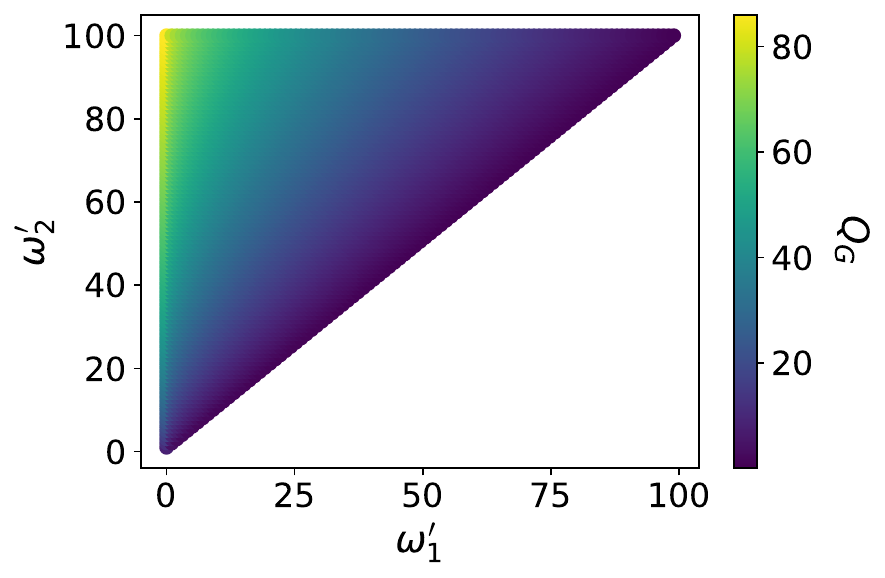}
        \label{fig:Q_G}
      \end{minipage} &
      \begin{minipage}[t]{0.45\hsize}
        \centering
        \subcaption{}
        \includegraphics[keepaspectratio, scale=0.4]{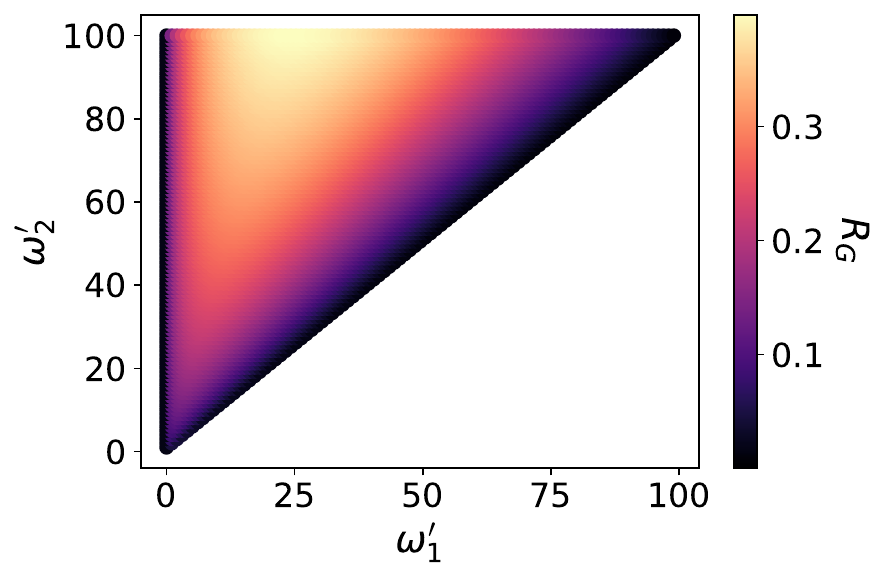}
        \label{fig:R_G}
      \end{minipage} \\
      \begin{minipage}[t]{0.45\hsize}
        \centering
        \subcaption{}
        \includegraphics[keepaspectratio, scale=0.4]{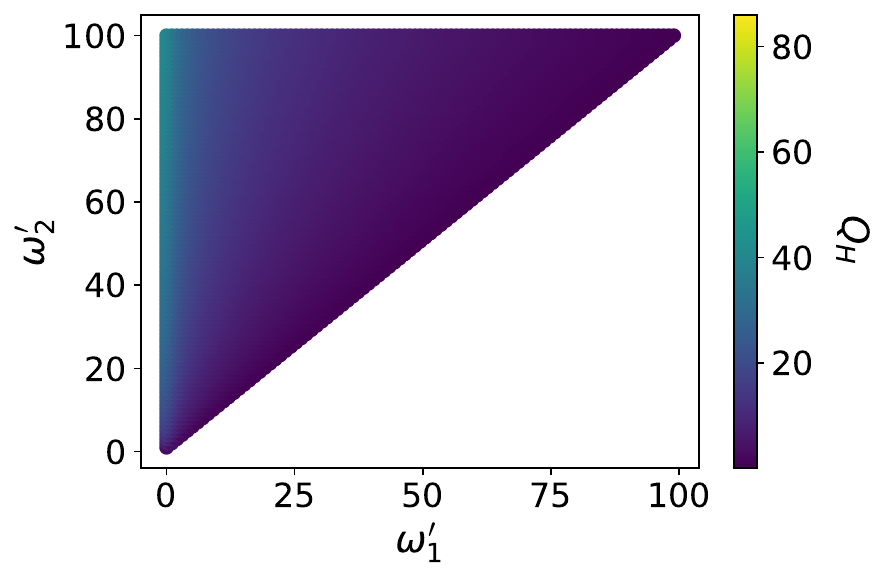}
        \label{fig:Q_H}
      \end{minipage} &
      \begin{minipage}[t]{0.45\hsize}
        \centering
        \subcaption{}
        \includegraphics[keepaspectratio, scale=0.4]{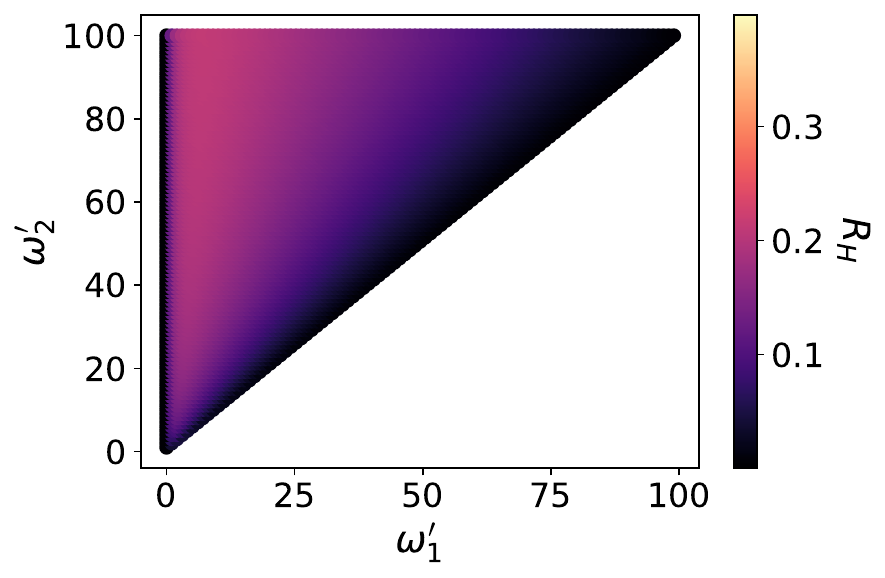}
        \label{fig:R_H}
      \end{minipage} 
    \end{tabular}
     \caption{The distortion indices (a)$Q_G$, (b)$R_G$, (c)$Q_H$, (d)$R_H$ for the different normalized frequencies $\omega^{\prime}_1, \omega^{\prime}_2$ with the normalized distance $\lambda=1$.}
\end{figure}

\subsection{Discussion on biological MC channels in nature}
We discuss the role of MC channels in nature from the perspective of signal distortion. In particular, we explore the range of frequencies in which signals can be transmitted with small distortion in specific MC channels in nature based on the signal molecular species (diffusion coefficients). 
More specifically, since it is known that the higher the frequency bandwidth of input signal, the larger output signal distortion from Proposition1 and 2, we analyze up to which higher frequency bandwidth signal distortion can be suppressed below a threshold value when transmitting a signal with a frequency bandwidth of one decade.
\par
In this discussion, we assume that signaling molecules diffuse through infinite space. Most signaling molecules used in MC are of three types: autoinducers, neurotransmitters, and ions. Each signal molecular species differs in the environment in which it is used for MC. For example, autoinducers are used for MC between bacteria in quorum sensing mechanisms \cite{waters2005quorum}. Assuming communication within a bacterial colony, the communication distance between neighboring bacteria is assumed to be at most $x_r=10\,\si{\micro m}$. Using the $Q_H$ and $R_H$ values calculated by Proposition 2, we define $Q_H/10$ and $R_H/10$ as thresholds. Since the diffusion coefficient is $\mu = 83\,\si{\micro m^2/s}$ \citep{karig2018stochastic}, it can be seen that the MC channel in quorum sensing can transfer synthesized signals with a bandwidth of 1 decade with little distortion up to the high-frequency bandwidth of $[2.0\cdot10^{-2},2.0\cdot10^{-1}]\,\mathrm{rad/s}$. Although it does not seem possible to transfer fast signals without distortion, the fact that the period of molecular reactions in bacteria is approximately several tens of minutes \citep{Kolodrubetz1981,weber2011noise} suggests that such reaction information can be transferred with some accuracy. On the other hand, neurotransmitters are mainly used for MC in the synaptic cleft between neurons \citep{ORourke2012}. The diffusion coefficient of a neurotransmitter is $\mu = 500\,\si{\micro m^2/s}$ (\cite{Soldner2020}), and the communication distance in the synaptic cleft is $x_r = 2.5\cdot10^{-2}\,\si{\micro m}$ \citep{ORourke2012}. We obtain $[1.9\cdot10^{4}, 1.9\cdot10^{5}]\,\mathrm{rad/s}$ as the highest frequency bandwidth of one decade width over which the distortion can be kept below $Q_H/10$ and $R_H/10$. Since the nervous system needs to accurately transmit fast signals to its terminals, it is inferred that the MC channel transfers signals in the high frequency bandwidth with small distortion by increasing the diffusion coefficient and shortening the communication distance. Ions are secreted out of the cell often through ion channels and ion pumps and are transferred to their destination by diffusion. Since MC by ions is found in various environments, the communication distance $x_r$ cannot be uniquely determined. However, since the diffusion coefficient is very fast as $\mu=500$ -- $7000\,\si{\micro m^2/s}$ \citep{Soldner2020}, MC using ions can be applied to situations in which fast signals are transferred without distortion or over a long distance and is therefore considered to be highly versatile. DNA molecules are beneficial as signal molecules for MC to transfer information. The proposed method can be also applied to larger molecules such as DNA molecules as long as they diffuse according to Fick's law. However, since the diffusion coefficient of DNA molecules is as small as $\mu=0.1$ -- $2.0\,\si{\micro m^2/s}$ \cite{Robertson2006}, DNA molecules would not be applicable to MC with a long distance compared to autoinducers, neurotransmitters and ions.
The parameters and the frequency bands for each molecular species are summarized in Table \ref{tab:species}.
\par
\begin{table}[t]
\caption{Biological parameters for different molecular species}\label{tab:species}
\begin{tabular*}{\textwidth}{@{\extracolsep\fill}lcccccc}
\toprule
Molecules & 
Diffusion coefficient $\mu$ & 
MC distance $x_r$ & 
Frequency bandwidth $[\omega_1, \omega_2]$ & 
\\
\midrule
Autoinducers  & $83\,\si{\micro m^2/s}$ &  $10\,\si{\micro m}$ & $[2.0\cdot10^{-2}, 2.0\cdot10^{-1}]\,\mathrm{rad/s}$ \\
Neurotransmitters  & $500\,\si{\micro m^2/s}$ & $2.5\cdot10^{-2}\,\si{\micro m}$ & $[1.9\cdot10^{4},1.9\cdot10^{5}]\,\mathrm{rad/s}$ \\
Ions  & $500$--$7000\,\si{\micro m^2/s}$ &\diagbox[height=0.8\line]{\ }{\ } &\diagbox[height=0.8\line]{\ }{\ }\\ DNA molecules & $0.1$--$2.0\,\si{\micro m^2/s}$ &\diagbox[height=0.8\line]{\ }{\ } &\diagbox[height=0.8\line]{\ }{\ }\\
\botrule
\end{tabular*}
\end{table}

\section{Conclusion}
In this paper, we have proposed a method to analyze signal distortion caused by diffusion-based MC channels. Specifically, we have first classified distortion into amplitude distortion and delay distortion, and defined the indices for these two distortions by the gain characteristics and the phase delay. We have then analytically derived these indices for distortion caused by diffusion based on the transfer function obtained by the diffusion equation. Finally, we have demonstrated the design procedure of the MC channel that causes less signal distortion using the proposed indices. 

\backmatter

\bmhead{Acknowledgments}
This work was supported in part by JSPS KAKENHI Grant Numbers JP21H01355, JP22J10554, and JP23H00506.

\appendix

\section{Proof of Proposition 1}
\label{sec:prop1}
We first consider the logarithmic gain characteristics $g(\omega)$ and phase delay characteristics $\tau(\omega)$ for the diffusion system $G(s)$. Using Euler's formula, the transfer function (\ref{eq:difsys}) can be transformed as
\begin{equation}
    G(j\omega) = e^{-\sqrt{\frac{x_r^2\omega}{2\mu}}}\left(\cos\sqrt{\frac{x_r^2\omega}{2\mu}}-j\sin\sqrt{\frac{x_r^2\omega}{2\mu}}\right),
    \label{eq:Geu}
\end{equation}
where we use 
\begin{equation}
    \sqrt{j} = \frac{1}{\sqrt{2}} + j\frac{1}{\sqrt{2}}.
\end{equation}
Using Eq. (\ref{eq:Geu}), the gain characteristics $|G(j\omega)|$ and phase characteristics $\angle{G(j\omega)}$ are obtained as
\begin{align}
 |G(j\omega)| = e^{-\sqrt{\frac{x_r^2\omega}{2\mu}}}, \ \angle{G(j\omega)} = -\sqrt{\frac{x_r^2\omega}{2\mu}}.
 \label{eq:absd}
\end{align}
Therefore, the logarithmic gain characteristics $g_G(\omega)$ and the phase delay characteristics $\tau_G(\omega)$ are
\begin{equation}
 g_G(\omega) = -20\sqrt{\frac{x_r^2\omega}{2\mu}}\log_{10}e,
 \label{eq:gaind}
\end{equation}
\begin{equation}
 \tau_G(\omega) = - \frac{\angle{G(j\omega)}}{\omega} = \sqrt{\frac{x_r^2}{2\mu\omega}}.
 \label{eq:taud}
\end{equation}
Since it is obvious that the logarithmic gain characteristics (\ref{eq:gaind}) and the phase delay (\ref{eq:taud}) monotonically decrease respect to frequency $\omega$, we have

\begin{align}
    \max_{\omega_1\leq\omega\leq\omega_2}g_G(\omega) = g_G(\omega_1), \ \min_{\omega_1\leq\omega\leq\omega_2}g_G(\omega) = g_G(\omega_2),
    \label{eq:gain1}\\
    \max_{\omega_1\leq\omega\leq\omega_2}{\tau_G(\omega)} = \tau_G(\omega_1), \ \min_{\omega_1\leq\omega\leq\omega_2}{\tau_G(\omega)} = \tau_G(\omega_2),
    \label{eq:tau2}
\end{align}
for $\omega > 0$. Using Eqs. (\ref{eq:gaind})--(\ref{eq:tau2}), we obtain the amplitude distortion $Q_G$ in Eq. (\ref{eq:Qd}) and the delay distortion $R_G$ in Eq. (\ref{eq:Rd}).

\section{Proof of Proposition 2}
\label{sec:prop2}

We first obtain the logarithmic gain characteristics $g_H(\omega)$ and phase delay characteristics $\tau_H(\omega)$ for the reception system $H(s)$. Substituting $j\omega$ into the transfer function Eq. (\ref{eq:H(s)}), the absolute value $|H(j\omega)|$ and the phase characteristics $\angle H(j\omega)$ are expressed as
\begin{equation}
 |H(j\omega)| = \frac{k_fr}{\sqrt{\omega^2 + {k_r}^2}},
 \label{eq:|H|}
\end{equation}
\begin{equation}
 \angle H(j\omega) = \tan^{-1}\left(-\frac{\omega}{k_r}\right).
 \label{eq:angleH}
\end{equation}
From Eqs. (\ref{eq:|H|}) and (\ref{eq:angleH}), we obtain the gain characteristics $g_H(\omega)$ and the phase delay characteristics $\tau_H(\omega)$ as
\begin{equation}
 g_H(\omega) = 20\log_{10}k_fr - 20\log_{10}\sqrt{\omega^2+{k_r}^2},
 \label{eq:reception_gain}
\end{equation}
\begin{equation}
 \tau_H(\omega) = -\frac{\angle H(j\omega)}{\omega} = \frac{\tan^{-1}{(\frac{\omega}{k_r})}}{\omega}.
 \label{eq:reception_phase_delay}
\end{equation}
From Eq. (\ref{eq:reception_gain}), $g_H(\omega)$ is obviously monotonically decreasing for $\omega>0$. Next, we show that $\tau_H(\omega)$ is monotonically decreasing with respect to $\omega>0$. The derivative of $\tau_H(\omega)$ for $\omega$ can be written as
\begin{equation}
   \frac{d\tau_H(\omega)}{d\omega} = \frac{1}{\omega^{2}} f(\omega),
   \label{eq:derivative_phase_recept}
\end{equation}
where 
\begin{equation}
  f(\omega) = \frac{k_r\omega}{k_r^{2} + \omega^{2}} - \tan^{-1}\left(\frac{\omega}{k_r}\right).
 \label{eq:omega_dash1}
\end{equation}
Since the derivative of $f(\omega)$,
\begin{equation}
  \frac{df(\omega)}{d\omega} =-2k_r\left(\frac{\omega}{k_r^{2}+\omega^{2}}\right)^{2},
  \label{eq:omega_dash2}
\end{equation}
is monotonically decreasing for $\omega>0$, and $f(0)=0$ holds, $f(\omega)<0$ for $\omega>0$. Thus, Eq. (\ref{eq:derivative_phase_recept}) is negative for $\omega>0$, which leads to that $\tau_H(\omega)$ is monotonically decreasing for $\omega>0$. Therefore, we have
\begin{align}
     \max_{\omega_1\leq\omega\leq\omega_2}g_H(\omega) = g_H(\omega_1), \ \min_{\omega_1\leq\omega\leq\omega_2}g_H(\omega) = g_H(\omega_2),
     \label{eq:gain3}\\
     \max_{\omega_1\leq\omega\leq\omega_2}{\tau_H(\omega)} = \tau_H(\omega_1), \ \min_{\omega_1\leq\omega\leq\omega_2}{\tau_H(\omega)} = \tau_H(\omega_2),
     \label{eq:tau4}
\end{align}
for $\omega>0$. Using Eqs. (\ref{eq:reception_gain}), (\ref{eq:reception_phase_delay}), (\ref{eq:gain3}) and (\ref{eq:tau4}), we obtain the indices shown as Eqs. (\ref{eq:Q_reception}) and (\ref{eq:R_reception}).


\end{document}